\documentclass{nature}
\usepackage{caption}

\usepackage{enumitem}
\usepackage{graphicx}

\usepackage{nameref}
\usepackage{xr-hyper}
\usepackage{hyperref}
\usepackage{textgreek}
\usepackage{lineno}

\usepackage[]{graphicx,bm,subfigure,amsmath}
\usepackage{placeins}

\makeatletter

\makeatletter

\usepackage[utf8]{inputenc}
\makeatletter
\let\saved@includegraphics\includegraphics
\AtBeginDocument{\let\includegraphics\saved@includegraphics}
\renewenvironment*{figure}{\@float{figure}}{\end@float}
\makeatother

\title{An ultrastrongly coupled single THz meta-atom\\}

\author{Shima Rajabali$^{1,*}$, Sergej Markmann$^{1}$, Elsa J\"ochl$^{1}$ , Mattias Beck$^{1}$, Christian A. Lehner $^{2}$ , Werner Wegscheider$^{2}$,  J{\'e}r{\^o}me Faist$^{1}$, Giacomo Scalari$^{1,*}$}

\usepackage[usenames, dvipsnames]{color}
\usepackage{soul}

\begin{document}

\maketitle
\begin{affiliations}
 \item Institute of Quantum Electronics, ETH Z{\"u}rich, 8093 Z{\"u}rich, Switzerland
 \item Laboratory for solid state physics, ETH Z{\"u}rich, 8093 Z{\"u}rich, Switzerland

\end{affiliations}
\begin{abstract}

 \textbf{
 Free-space coupling to strongly subwavelength individual optical elements is a central theme in quantum optics, as it allows to control and manipulate the properties of quantum systems. 
In this work, we show that by combining an asymmetric immersion lens  setup and complementary design of metasurfaces we are able to perform THz time domain spectroscopy of an individual, strongly subwavelength ($\frac{d}{\lambda_0}=1/20)$ meta-atom. We unravel the linewidth dependence of planar metamaterials as a function of the meta-atom number indicating quenching of the Dicke superradiance.
 On these grounds, we investigate ultrastrongly coupled Landau polaritons at the single resonator level, measuring a normalized coupling ratio of $\frac{\Omega}{\omega}=0.60$ resulting from coupling of the fundamental mode to a few thousand electrons.
Similar measurements on a low loss, less doped two dimensional electron gas yield a coupling ratio $\frac{\Omega}{\omega}=0.33$ with a cooperativity $C=\frac{4g^2}{\kappa \gamma}= 94$. Interestingly, the coupling strength of a coupled single resonator is the same as of a coupled array.
Our findings pave the way towards the control of light-matter interaction in the ultrastrong coupling regime at the single electron/single resonator level. The proposed  technique is way more general and can be useful to characterize the complex conductivity of micron-sized samples in the THz and sub-THz domain.}
\end{abstract}

Extreme electronic and photonic confinement has recently allowed a number of groundbreaking advances in several fields, from fundamental studies \cite{ChikkaraddyNature:16} to applications \cite{IEEE_IGA_VCSEL_2000,SchullerNat_Mat2010}. Particularly interesting is the possibility to manipulate \textit{on-chip} the light matter-coupling achieving new quasi-particles called cavity polaritons \cite{KavokinBaumbergBook} that profoundly modify the electro-optical properties of the constituent elements.

Cavity polaritons rely on enhanced vacuum field fluctuations $\mathcal{E}_0$ to reach large values of the vacuum Rabi frequency $\Omega_R \propto \mathcal{E}_0$ quantifying large values of coupling \cite{Hagenmueller2010}. The dependence from the inverse of the cavity volume $\mathcal{E}_0 \propto 1/\sqrt{V}$ led to the development of strongly subwavelength structures to reach the strongest couplings to-date \cite{BallariniNanophot2019,Forndiaz2019}. Metallic-based cavities have been used in the mid-infrared (Mid-IR)\cite{malerbaapl2016} and terahertz (THz)\cite{Todorov2010} regions of the electromagnetic spectrum for strong coupling experiments. Due to their subwavelength nature, the experiments have been carried out on multiple cavities ($100-1000$ or more) in order to enhance the signal-to-noise ratio (SNR). Since the frequency components excited for subwavelength features cannot propagate in the far-field, only recently  near-field techniques proved to be successful in probing single resonator-based  strongly coupled systems at Mid-IR frequencies at room temperature \cite{Gillibert2020APL,NanolettWangBrener2019}. 
 In the field of ultrastrong coupling, the development of the Landau polariton platform\cite{Scalari2012} has led to very fascinating results,  including studies of  non-linear phenomena \cite{LangePRL2021Nonlin},  the reaching of  high cooperativity \cite{KonoNatPhys2016} and the study of magnetotransport \cite{ParaviciniNatPhys2019} in ultrastrogly coupled Hall systems \cite{Felice_Transp_2021ARXIV}. In this approach,  extremely subwavelength planar split-ring resonators have been used in combination with one-dimensionally confined semiconductor heterostructures to achieve normalized coupling values $\frac{\Omega}{\omega}>1.4$ (Ref \cite{Bayer2017}). 

The light-matter interaction strength in most of the solid-state-based cavity quantum electrodynamics (QED)  systems scales with $\sqrt N$ with $N$ electron number due to the collective (Dicke) enhancement of the interaction strength\cite{Dicke1954}. In  Landau polaritons,  when employing metasurface cavities, the number of coupled electrons is shared among several hundreds of identical cavities. In the last few years, there has been a growing interest in studying ultrastrong light-matter coupled systems towards the limit of few electrons coupled to a single cavity \cite{TodorovPhysRevX,Keller2017,Jeannin2020} to study the fermionic Rabi model for the coupled systems rather than the bosonic Hopfield description where both material excitation and the electromagnetic field
are described as boson fields. Moreover, such systems can be building blocks for performing quantum information tasks.
In this perspective a reduction in the number of coupled carriers towards the low limit of a few and ultimately a single coupled element constitutes a great challenge  both in terms of measurement sensitivity \cite{ChikkaraddyNature:16,HaleLPR2020} and optical cavity design.

The possibility and the limits on the interaction of free propagating far-field optical beams from individual elements (molecules, atoms, quantum dots etc... ) have been investigated both theoretically \cite{PhysRevLett.101.180404AGIO,ModeconvEurphysLett2007} and experimentally \cite{Pinotsi_PhysRevLett.100.093603 ,NanolettImmersQDOTIma2007}, mainly at visible or infrared frequencies. 
In our case the individual element is not a quantum object but rather a single-subwavelength metallic planar resonator, operating at millimeter wavelengths, lying in-between free space propagating beams and guided microwaves. The single resonator spectroscopy at sub-THz/THz spectrum is performed by implementing a simple and practical asymmetric Silicon immersion lens (aSIL) configuration. We also demonstrate the linear dependence of the cold cavity linewidth of a 2D array of the metasurface on the number of resonators in the array when the illuminated area with resonators becomes comparable or smaller than the wavelength. This observation is an indication of quenching the superradiance decay. Moreover, the Landau polaritons in a coupled single-subwavelength resonator are resolved on two different semiconductor heterostructures, and high normalized coupling strength up to $60\%$ and high cooperativities up to $C = 94$ are achieved.

\section*{Measurements and results}
\subsection{Single resonator detection and superradiance}
\begin{figure}
    \centering
    \includegraphics[width=1 \textwidth]{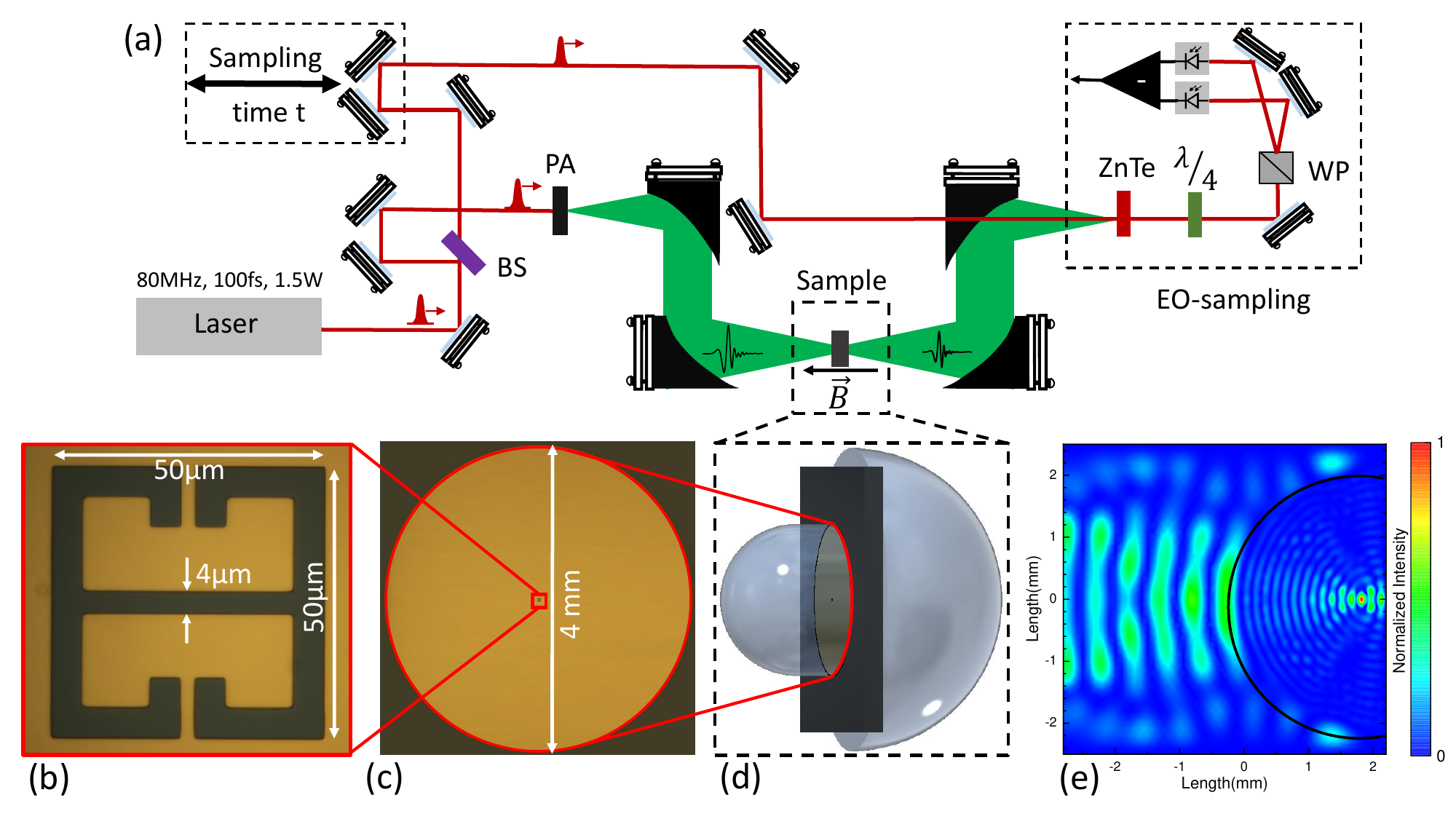}
    \caption{\textbf{Schematic of the TDS setup with the aSIL configuration} (a) The schematic sketch of the THz-TDS used for the transmission experiments. (b) The optical image of the single resonator including the periodicity and gap size. (c) The optical image of the whole sample metallized with a circular boundary of diameter $4 mm$ for the alignment of the front lens. (d) The magnified schematic of the aSIL configuration. (e) The COMSOL simulation result of the focused THz beam at the interface between the front lens and the resonator field.}
    \label{fig:Schematic}
\end{figure}

Free-space probing of a sample containing a few/single subwavelength resonators  is difficult as it is intrinsically inefficient in exciting the right electromagnetic modes and generally features a very low signal-to-noise ratio. 
We thus developed a strategy based on the employment of the complementary metamaterial approach combined with  an aSIL configuration to optimize the matching of a free space propagating beam to a single element of a complementary metasurface. 
The conventional THz-time domain spectroscopy (TDS) setup  has a pair of off-axis parabolic mirrors first to collimate and then focus the incident THz beam from the photo-conductive switch\cite{MadeoAntennaEL2010} on the sample. The transmitted signal from the sample will be then collected, collimated, and then focused on the ZnTe crystal through another pair of off-axis parabolic mirrors (Fig.~\ref{fig:Schematic}a). The detection is performed using an electro-optic detection scheme\cite{Gallot1999}. The beam, that we can assume has a Gaussian profile, in this setup has a spot size of around $3 mm$ at the sample's location. 
In order to correctly match the incoming THz beam to the single complementary meta-atom, we designed a back-to-back lens system inspired by some previous studies on single quantum dots at visible wavelengths\cite{ NanolettImmersQDOTIma2007} (Fig.~\ref{fig:Schematic}d). Being very akin to a $4f$ arrangement, a similar setup has been recently proposed also as a THz spatial filter \cite{Gan2020}. A set of ray optics simulations in COMSOL Multiphysics is done to find the correct lens dimensions and adjust the input and output numerical aperture of the system such that it is compatible with the existing THz-TDS setup. A final design consists of a front hyperhemispherical Silicon (Si) lens of diameter $4 mm$ that focuses the incident beam to a spot size of $\sim 500 \mu m$ at the surface of the sample (Fig.~\ref{fig:Schematic}e), where the metamaterial resonators are located. 
The back hemispherical Si lens of diameter $8 mm$ collects the transmitted beam from the sample and sends it through the off-axis parabolic mirrors to the detector. To optimize the dimensions for the back lens, we run a parametric sweep on the back lens diameter in the COMSOL simulation. The goal was to match the output numerical aperture of the lens system with the numerical aperture $f\#3.85$ of the off-axis parabolic mirror according to the chosen thickness and material for the substrate at the operating frequency ($f = 300 GHz$). In our case, we used a $500 \mu m$ thick semi-insulating gallium arsenide (GaAs) slab that corresponds to the average thickness of the substrates used for the epitaxial growth of the samples. The presence of the substrate where the sample is deposited also makes the back lens effectively an hyperhemispherical one.  
The significant advantage of this asymmetric design compared to the previous symmetric designs\cite{Gan2020, NanolettImmersQDOTIma2007} is that there is no need to sandwich the target surface (in our case, the metamaterial metallic thin film) between two semiconducting substrates to have the surface at the focus point of the confocal system. The asymmetry of the lenses accounts for the sample substrate thickness as an active part of the optics. The sample itself is located on top of the substrate and has a very small thickness compared to the wavelength. The planar complementary metasurface and the underlying quantum well (QW) sum up to less than $1 \mu m$ in the propagation direction of the THz beam.
Thus the back lens with a larger diameter compensates for the substrate thickness and collects the diverging beam at the back interface of the substrate. The performance of our aSIL setup can in any case be further improved by a careful choice of the lens curvature to optimize the matching with the incoming THz beam and by providing the Si lenses with a broadband anti-reflection coating to further enhance the SNR \cite{Gan2020}.

 \begin{figure}
    \centering
    \includegraphics[width=1 \textwidth]{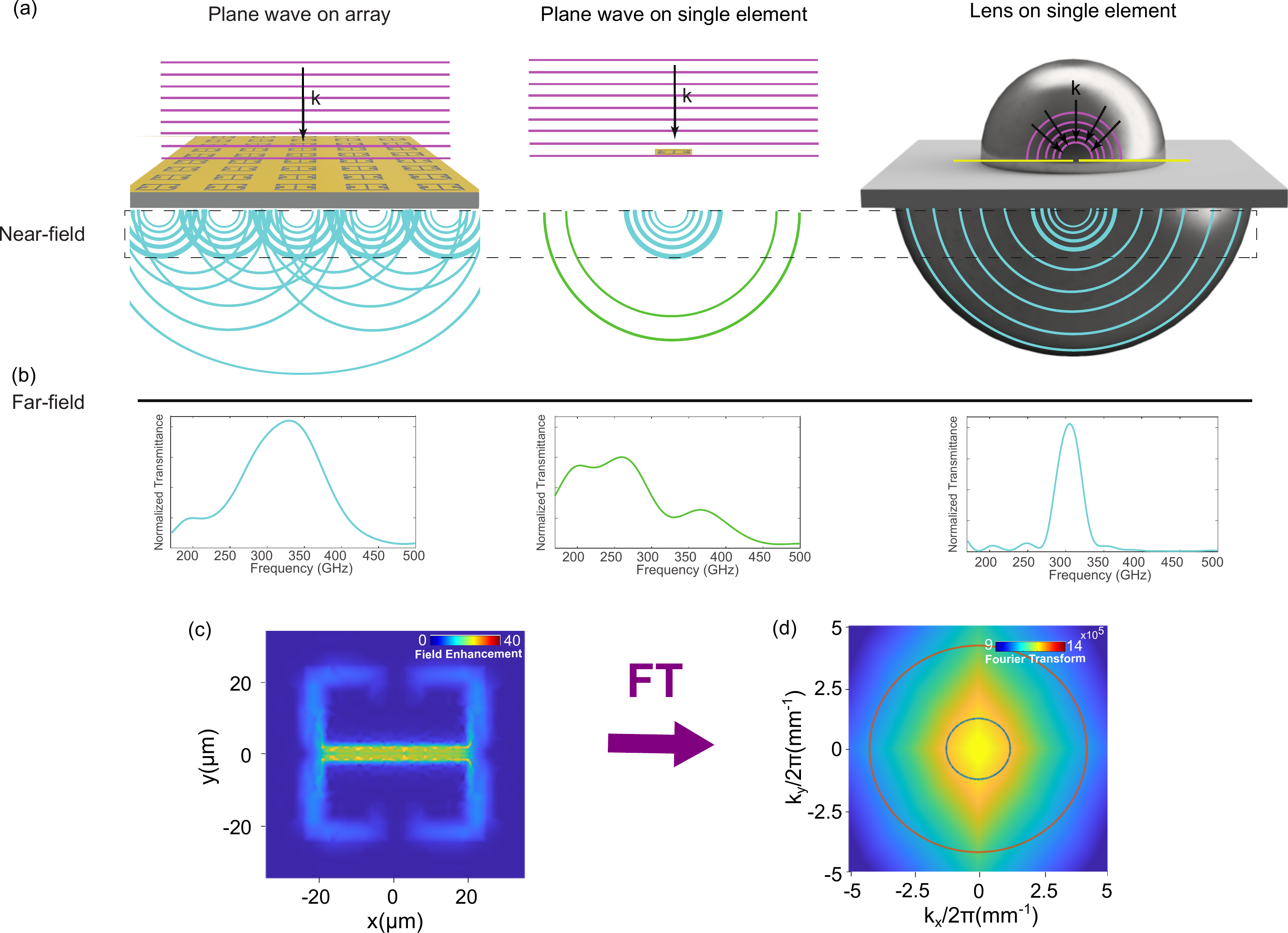}
    \caption{\textbf{No lens vs. aSIL configuration for Single resonator spectroscopy}(a), (b): A schematic of the near-field (a) and far-field (b) of an array without lenses and a single resonator in case of without and with aSIL configuration. (c) The finite element simulation result of the field enhancement at the resonator plane for a single resonator with an incident plane wave with a polarization along y-axis. (d) 2D Fourier transform of the field in panel (c). The blue and red circles represent the accessible wave vectors for far-field propagation in case of having incident waves in the air (without lenses) and in Si(with the aSIL arrangement), respectively.}
    \label{fig:metamatscheme}
\end{figure}

To-date, single meta-atom spectroscopy  has been possible only with near-field techniques \cite{HaleLPR2020,Gillibert2020APL,NanolettWangBrener2019}. 
Here we employ the back-to-back aSIL setup to study a single complementary resonator. We previously demonstrated the efficiency of the complementary THz metasurfaces in the context of Landau polaritons\cite{MaissenPhysRevB2014}. The significant advantage of this configuration in the transmission measurement of a single resonator is that the measured signal above the noise level is directly the signal of interest yielding a very high contrast. In the case of a standard or direct metasurface (highly transmissive sample),  the challenge of the measurement is to detect a very small absorption change on top of a large signal. If one employs a complementary metasurface, the situation is reversed as the transmitted signal is very low, but the only signal detected is the relevant one. Fig.~\ref{fig:metamatscheme} visualizes the reasons that single resonator spectroscopy is possible using an aSIL configuration. The front and back hyperhemispherical Si lenses with a refractive index of $n_r = 3.42$ at $300 GHz$ \cite{Grischkowsky1990} focuses the beam onto the resonator plane. 
The front lens enhances the intensity of the beam arriving at the resonator plane  (Fig.~\ref{fig:metamatscheme}a) and both lenses expand the accessible wave vectors for far-field propagation (Fig.~\ref{fig:metamatscheme}c). The back lens improves the collection of the signal and increases the frequency resolution due to impedance matching and suppression of the echos from the interface between back of the sample and the back lens.
\par In Fig.\ref{fig:metamatscheme} we clarify, using a Fourier optics argument,  the mechanism at the basis of the different signals observed when probing a metasurface with a large number of resonator and a single meta-atom  with a plane wave compared to the case of a single meta-atom illuminated with an immersion lens. 
The meta-atoms constituting the planar metamaterial are strongly subwavelength by design: they can be usually probed by far-field optics because the near-field components diffracted by each element are scattered in the far-field by the surrounding elements of the array. The unit cell dimension d$_{unit}$ is highly subwavelength as well (in our case d$_{unit}=70 $ $\mu$m for a free space wavelength $\lambda_{fs} \simeq$ 1 mm, with $\frac{d_{unit}}{\lambda_{fs}} = 7 \times 10^{-2}$ ) allowing the spatial frequencies to propagate (Fig.~\ref{fig:metamatscheme}a and \ref{fig:metamatscheme}b). The presence of the resonator array is as well required for the excitation of the resonant mode as the nearby elements diffract the incoming wave providing some in-plane components of the wave vectors.   In the case where the plane wave illuminates a single subwavelength meta-atom  (Fig.~\ref{fig:metamatscheme}c) a large portion of spatial frequencies remain trapped in the resonator's near field and do not reach the far-field detector. The corresponding spectrum does not show the resonant feature at 300 GHz. When the aSIL arrangement is used (Fig.~\ref{fig:metamatscheme}d), the spread in the wave vector excites the correct resonator mode that can be then collected from the back lens and recover the searched transmission resonance. In the lower row of the figure we plot the near-field electric field distribution and its spatial Fourier transform. The circles correspond to the wave vectors propagating in the case of without lenses (blue) and with lenses (red). It is evident that the quasi totality of wave vectors can propagate in the far-field. As a further confirmation of our model, we can cite the results obtained with a near-field probe measuring a similar complementary single resonator \cite{HaleLPR2020}. In that experiment, the single resonator was excited with a focusing lens, providing the necessary in-plane k's. The resonant signal was found to remain trapped in the near field since there was no collecting lens. 
Now that we clarified the mechanism that underlies the observed transmission spectra, we can discuss the experimental measurements in detail. 

As a first step, we perform a systematic study of the cold cavity, illuminating a series of samples deposited onto a semi-insulating GaAs substrate. These samples contain a decreasing amount of resonators, from $3600$ down to a single one. We study the linewidth and the transmission amplitude of the cold cavity and their dependency on the number of resonators. 
The theory of the superradiance \cite{Dicke1954,Keller2018,Sersic2009,Jenkins2012} predicts a reduction in the collective radiative decay and consequently an improvement in quality factor by reducing the number of two-level emitters confined  in a spatial region comparable or smaller than  the wavelength. This has also been investigated experimentally on metamaterials but never in the single resonator limit  \cite{Choudhary_PRA_2019}. In order to study this effect, we designed and fabricated 2D array of $n\times n$  complementary split-ring resonators (cSRR), including a very large array of 60 $\times$ 60 resonators and small arrays with a varying $n$ from 6 to 1, operating at a resonant frequency of 350 GHz and a periodicity of 70 $\mu m$ on a GaAs substrate (Fig. \ref{fig:Schematic}b).

\begin{figure}[h!]
    \centering
    \includegraphics[width=1 \textwidth]{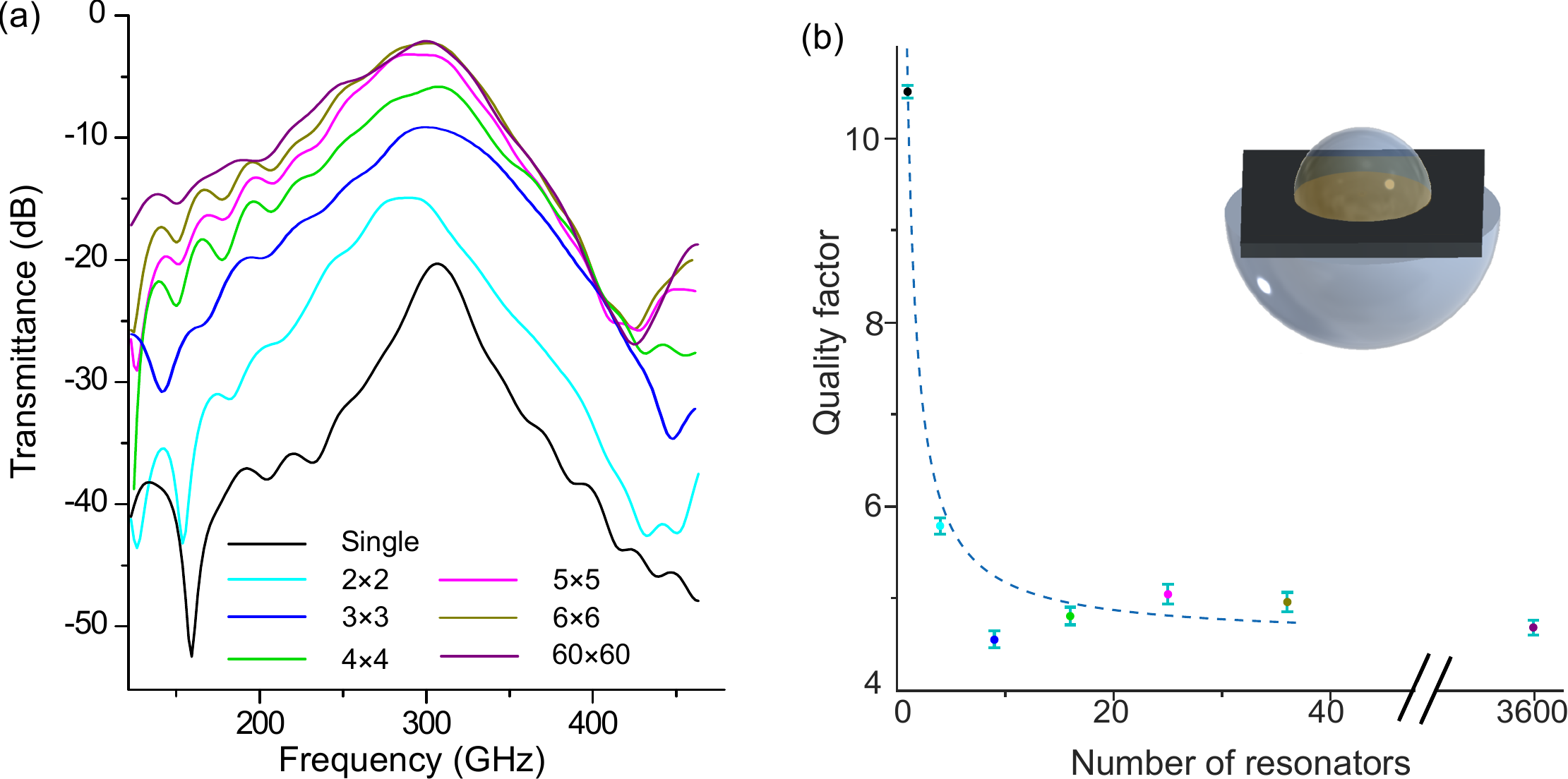}
    \caption{\textbf{Quenching of the superradiance decay} (a) Transmittance of $n \times n$ array of cSRRs ($n = 1$ to $6$) and a large array with $n = 60$. These measurements are done using the aSIL assembly  and the resonant peak of the single resonator is resolved. (b) is Q factor 
    vs. the number of resonators extracted from the measurements with lenses. The color of the data points are the same as the color of the transmittance curves in panel (a). The blue dashed line shows the linear dependence of the quality factor on the density of the resonators ($Q = \frac{6.04}{N}+4.6$). The Q-factors are calculated using fitting the resonant peak with BWF function. All error bars represent one $\sigma$ confidence interval.}
    \label{fig:Menlo_meas}
\end{figure}

We measured the resonators with a commercial THz spectrometer (details in Methods section) at room temperature. We conducted the measurements with and without the back-to-back aSIL system . As expected\cite{HaleLPR2020}, the resonant peak of  $2 \times 2$ cSRR array and the single resonator could not be resolved without the use of our aSIL arrangement  because of the weak interaction of a few resonator with the THz beam (see the supplementary info). Even the measurement with $9$ resonators shows a very weak resonant peak. 
In Fig. \ref{fig:Menlo_meas}, we report the result of the study, showing the collected spectra using the aSIL assembly and the extracted quality factors (Q-factor) for measurement with lenses. 
The measurement shows about 5 times signal enhancement in the time domain for the single resonator. Moreover, the resonant peak for the arrays with less than $9$ resonators were resolved with a large dynamic range ($60 dB$ for the single resonator).

Another key feature of this technique is the significant improvement in frequency resolution. Since Si and GaAs have very close refractive indexes up to $1 THz$ \cite{Grischkowsky1990}, the echoes of the THz pulse, reflected at each interface are greatly suppressed due to the impedance matching and, as a consequence, the etalon effects are minimized. Particular care was taken to assemble the stack lens-sample-lens without any air gap, pressing on the assembly with metallic positioners (see Supplementary material).
In the best conditions, the resolution was $18 GHz$ (corresponding to a time scan length of $55 ps$), limited by the reflection from the detector.
If we compare the results of the measurements on the large arrays with and without the lenses we observe a difference in the resonant frequency due to the presence of the front Si lens. The presence of an high index material in the near field of the complementary resonators  red-shifts the resonance to values below $300 GHz$ which is also confirmed by our finite element simulations (more information in the supplementary document).
\par To calculate the Q factor, we fitted the resonant peak with a Breit-Wigner-Fano (BWF) function. Fig.\ref{fig:Menlo_meas} shows an almost constant Q-factor for $3600$ to $16$ resonators measured with lenses. For the samples with a fewer number of resonators ($n \leq 3$), as the dimension of the resonator field becomes comparable to the wavelength (at $300GHz$, the wavelength propagating in GaAs, $\lambda_n=\lambda/n_{GaAs} =278 \mu m$ and the one propagating in Si $\lambda_n=\lambda/n_{Si} = 292\mu m$), the quenching of superradiance emerges. The Q-factor doubles from $9$ resonators to a single resonator due to the quenching of the superradiant decay and reaches $\sim$ 11 for the single one. This can have important consequences in the context of the strong coupling experiments that we will discuss in the next section.  
If we compare our results with similar studies performed on arrays of metamaterials down to a few elements \cite{FedotovPRL2010}, we observe a similar behaviour as in the case of "incoherent" resonators that are not coupled by magnetic interaction, since the symmetric arrangement of the inductors in our case cancels out the magnetic response. What is clear in our data is the enhancement of the Q factor when the occupied area by the resonators is smaller than $\lambda ^2$. 
It is worth noting that the number of illuminated resonators in case of the large array ($60 \times 60$ resonator) are 2628  for the measurement without lenses (a $4 mm$ diameter illuminating spot) and 44 for the measurement with aSIL configuration (a $500 \mu m$ diameter illuminating spot).

 \subsection{Ultrastrong coupling in a single complementary resonator} 

\begin{figure}
    \centering
    \includegraphics[width=1 \textwidth]{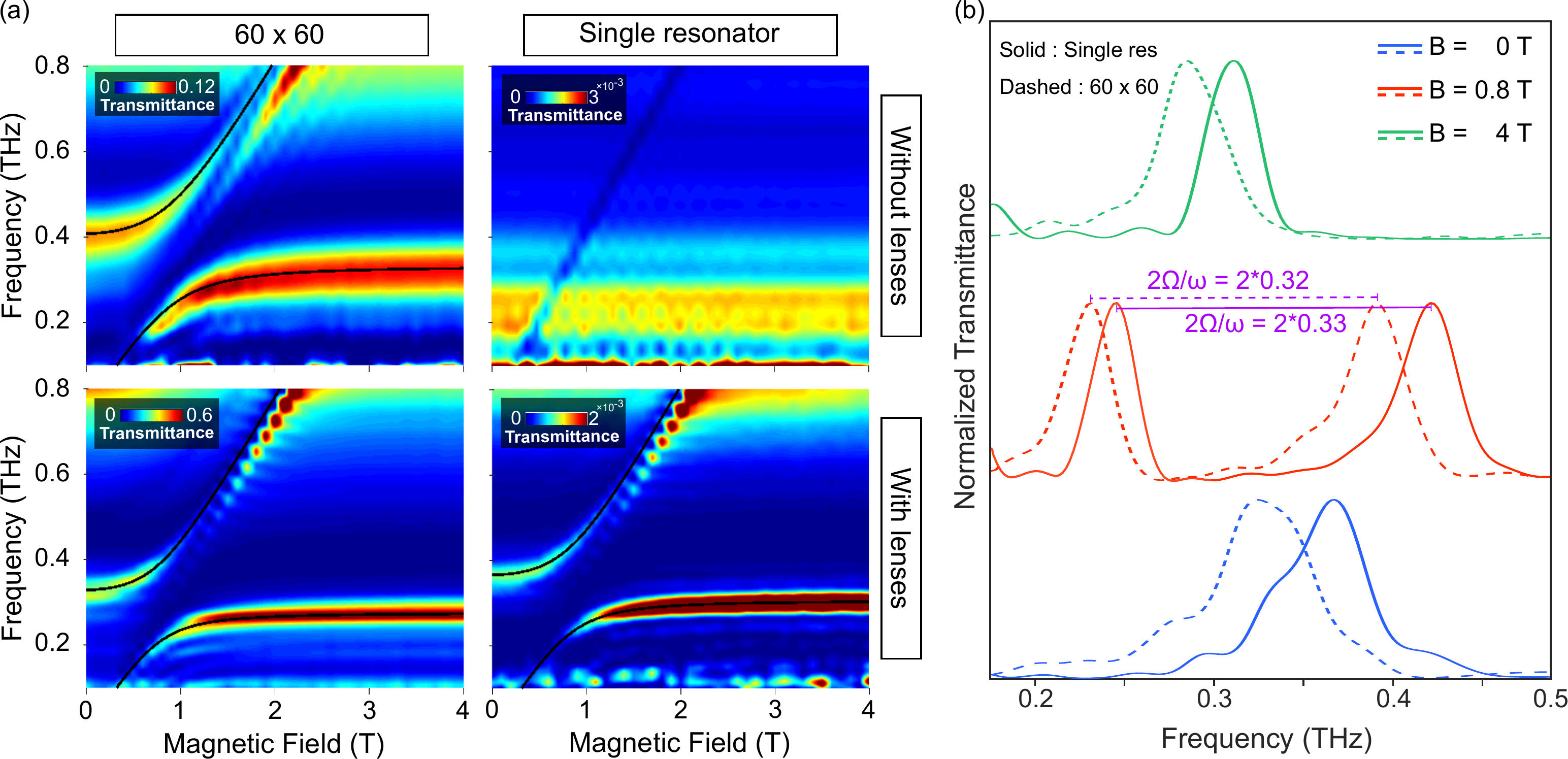}
    \caption{\textbf{Comparison of THz-TDS measurements of a large array and a single cSRR coupled to LL transitions in a GaAs QW with/without Si lenses}(a)  The transmission measurements without (top row) and with the aSIL configuration (bottom row) for a 2D array of $60 \times 60$ cSRRs (left column) and a single resonator (right column). The polariton branches in the measurement of the single resonator  without lenses are not resolved due to the weak interaction of the resonator with the THz beam. The black solid lines are the fitted LP and UP using the Hopfield model. (b) Sections of the colormap for the single resonator (solid lines) and $60 \times 60$ array () both measured with lenses at different magnetic field values ($B = 0,  0.8 T,$ (anti-crossing) and $4 T$) to demonstrate the well-resolved polariton branches. The distance between peaks at $B = 0.8 T$ which is twice of the vacuum Rabi splitting is marked with purple. All the modes are normalized to their maximum value. The vertical shift is for clarity.}
    \label{fig:Oxford_GaAs}
\end{figure}

After having characterized the measurement technique and studied the Q-factor dependence of the cSRR array on the number of resonators, we apply the developed method to the spectroscopy of a single resonator ultrastrongly coupled to an inter-Landau level (LL) transitions in a single GaAs QW. With a similar fabrication process, a large array and a single cSRR were deposited on top of a two dimensional electron gas (2DEG) produced in a single GaAs square QW located $90 nm$ below the surface. The sample transmission is then measured in a THz-TDS setup at cryogenic temperature ($T = 2.7 K$) as a function of magnetic field swept between $0$ and $4 T$. The best resolution for this setup is $33 GHz$ (corresponding to a time scan length of $30 ps$), limited by the thickness of the cryostat windows. Further details are to be found in the Methods section. 
In order to gain insight into the coupling of the complete metasurface and the single resonator to the THz beam, we measured the transmission spectra for both kind of samples with and without the aSIL assembly. The results, displayed as transmittance color maps as a function of the magnetic field, are reported in  Fig.~\ref{fig:Oxford_GaAs}. The figure shows a comparison between the transmission measurements without (top row) and with aSIL configuration (bottom row) for a 2D array of $60 \times 60$ resonators (left column) and a single resonator (right column). The colormaps relative to the plane wave (without lenses) case clearly show very well resolved polaritonic branches for the $3600$  resonator sample and a very different spectrum  in the case of a single resonator.
In the single resonator measurement, we observe a broad spectral feature corresponding to high transmission that extends from $180 GHz$ to $300 GHz$ and is independent from the applied magnetic field. A narrow absorption feature linearly changing with the magnetic field corresponding to the cyclotron resonance is crossing the broad transmission peak.
The single resonator excited by a plane wave, whose radiation is collected without an immersion lens, is not showing ultrastrong coupling since the correct resonant LC mode cannot be excited and detected. 
The measurement in the case of the single resonator measured with the aSIL assembly is significantly different: we observe extremely well resolved polaritonic branches that compare well with the one observed in the $3600$ resonator sample in both cases with and without the lenses. 
We can now analyze in detail and compare the polaritons measured in the case of employing the aSIL assembly for the single resonator and the large array.

The lower polariton (LP) mode of the coupled single resonator at its asymptotic limit, at $B = 4 T$, has a Q-factor of $15.4$ (using time trace decay method). The Q-factor of the LP  at $B =4 T$ for the sample with the coupled 2D array of $3600$ resonator without lenses and with lenses are $3.4$ and $6.9$, respectively. Similar to the cold cavity in the previous section, the Q-factor of the LP at its asymptotic limit is higher for the single resonator compared to the one for the array due to the quenching of the superradiance decay. By extracting the maximum of the spectrum at each magnetic field and fitting them with Hopfield model\cite{Hagenmueller2010}, a normalized coupling of $\frac{\Omega}{\omega}= 33\%$  is achieved for the single resonator measured in a confocal configuration. For the array of $3600$ resonators the normalized coupling for the measurement with and without lenses are $\frac{\Omega}{\omega}= 32\%$ and $36\%$, respectively. 
The cross sections of the colormap of the $60 \times 60$ array and the single resonator measurement with lenses at three different magnetic field values of $0T, 800 mT,$ and $4T$ show the well resolved upper polariton (UP) and LP peaks in the single resonator measurement (Fig.~\ref{fig:Oxford_GaAs}b). We notice that, as expected, the polaritonic branches in the single resonator case display a higher quality factor with respect to the 60x60 array. 
It is also evident that there is a renormalization of the loaded cavity frequency that blueshifts the resonance by 30 GHz ($\simeq 0.1 \omega$) when we reduce the number of resonators from 3600 to 1.  Interestingly, the normalized coupling strength remains basically the same ($\frac{\Omega}{\omega}\vert_{single}=0.33$, $\frac{\Omega}{\omega}\vert_{3600}=0.32$). We do not observe such a shift when we investigate the cold cavity at 300 K (see Fig.\ref{fig:Menlo_meas}(a)). We attribute this shift to the large dielectric contribution of the 2DEG that effectively produces a slow light effect enhancing the collective Lamb shift of the ensemble of meta atoms as observed in other systems \cite{Yao_metamatPhysRevB_2009, Frucci_2017,Wen_PhysRevLett2019}.

The calculated cooperativity for the coupled single resonator measurement is equal to $C = 94$ and for the large array of resonators are $C = 26.4$ without lenses and $C = 37$ with the aSIL configuration.

The cooperativity is calculated using $C = \frac{4g^2}{\kappa\gamma}$ where g is half of the mode splitting at the anti-crossing, $\kappa$ is the mode dissipation (cavity decay) rate and $\gamma$ is the decoherence rate. $\gamma$ is extracted from the direct measurement of the cyclotron coherence time \cite{Wang2010} using TDS measurement of the 2DEG measured with the confocal system. The linewidth of the cyclotron resonance is smaller than the one measured without lenses due to the superradiance effect (more information in the supplementary document). An estimate of the number of optically active electrons at the top most LL for our single cavity with a cavity surface of $S = 155 \mu m^2$ on a single GaAs QW yields $n = \frac{eB}{\hbar}\times S = 2.42 \times 10^{14} [\frac{1}{m^2T}] \times B \times S \simeq 30000$ (Ref \cite{Keller2017}).

\par
\begin{figure}
    \centering
    \includegraphics[width=1 \textwidth]{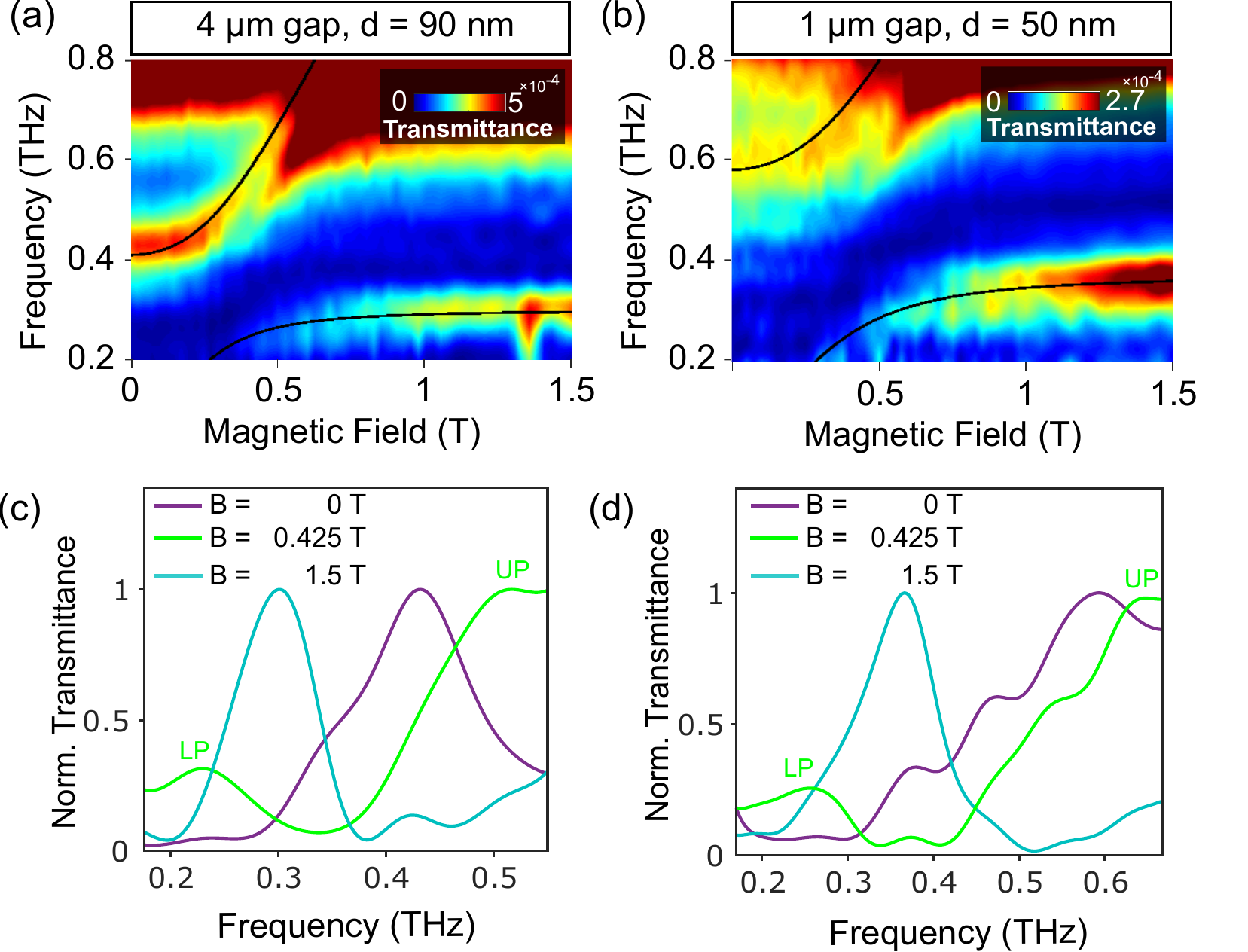}
    \caption{\textbf{ THz-TDS measurements of  LL transitions in a single InSb QW coupled to a single resonator} (a) and  (b): The colormaps correspond to the resonators with a gap size of 4 $\mu m$ on a InSb QW located at a distance of 90 nm below the surface (a) and a gap size of 1 $\mu m$ on a shallower InSb QW located at a distance of 50 nm below the surface (b), measured with lenses. They show normalized coupling ratio of $\frac{\Omega}{\omega}= 47\%$ and $60\%$ and cooperativity of $C = 12.5$ and  $14$, respectively. The black solid lines are the fitted LP and UP using the Hopfield model. (c) and (d): Sections of the colormaps in panel (a) and (b) are represented in panel (c) and (d), respectively, at different magnetic field values (B= $0, 0.425 T$ and $1.5T$). To have a more visible LP and UP, a section at B = $0.425 T$ which is at a marginally higher value than the anti-crossing is chosen. }
    \label{fig:Oxford_InSb}
\end{figure}

 In order to increase the coupling and reduce the number of optically active electrons, we employ an Indium Antimonide (InSb)-based 2DEG \cite{JKellerPRB2020nonparab,Lehner2018} at the same distance from the surface as the GaAs 2DEG (90 nm). Due to the lighter effective mass of electrons in InSb ($m_{eff} = 0.0225 m_0$), the anti-crossing for the coupled system lies at a much lower magnetic field value ( $250 mT$ compared to GaAs at $800 mT$) for the same resonator frequency of $300 GHz$ (Fig.~\ref{fig:Oxford_InSb}a). Therefore, the number of coupled electrons is reduced more than $3$ times with respect to the GaAs case. To further reduce the cavity volume and increase the field confinement, a single cSRR with a narrower gap of $1 \mu m$  at the same resonant frequency ($300 GHz$) is also fabricated and measured (Fig.~\ref{fig:Oxford_InSb}b). To enhance the coupling, this cavity is deposited onto a shallower InSb 2DEG with the same QW thickness at a distance of $50 nm$ below the surface. The cavity surface, evaluated with finite element simulations,  is $S = 138.32 \mu m^2$ 
for the larger gap and $S = 28.8 \mu m^2$ for the smaller gap. The number of coupled electrons at the anti-crossing are then $n_e = 8368$ and $n_e = 2090$ \cite{Keller2017} for the resonators with the gap size of $4 \mu m$ and $1 \mu m$, respectively.

If we extrapolate this result and consider the case of our previously demonstrated highly confined, hybrid dipole antenna split-ring resonator \cite{Keller2017} the use of an InSb QW can lead to an ultra-strongly coupled system with less than $5$ coupled electrons.

\par Due to a higher absorption in InSb QWs, the transmitted signal is less than in  the single resonator measurement on GaAs-based QW. The LP mode at its asymptotic limit (at B = 4T) has a Q-factor of 6.4 in both cases. The normalized coupling strengths are increased and equal to $\frac{\Omega}{\omega}= 47\%$ and $60\%$ for a single cSRR with 4 $\mu m$ and 1 $\mu m$ gap (more information about the fitting in the supplementary document). The cooperativity in the case of these InSb based samples are $C_{gapsize:4\mu m} = 12.5$ and $C_{gapsize: 1\mu m} = 14$. The lower cooperativity (compared to the values for a single resonator on a single GaAs QW) despite having a higher coupling strength is due to an order of magnitude lower mobility in InSb QWs which affects the decoherence rate (larger $\gamma$).\\
 The cross sections of the colormaps of the single resonator on InSb QWs measured with lenses at three different magnetic field values of $0 T, 425 mT,$ and $1.5 T$ show the well resolved upper polariton (UP) and LP peaks in the single resonator measurement (Fig.~\ref{fig:Oxford_GaAs}c , \ref{fig:Oxford_GaAs}d). Since UP and LP are not both visible at the anti-crossing ($\sim 250 - 300 mT$), a cross section at a slightly higher value at $B = 425 mT$ is represented. It is worth noting that in the narrower gap resonator, the broadening of UP and partially disappearance of LP at low values of the magnetic field are ascribed to their coupling to a continuum of magnetoplasmon excitations as recently discussed in Ref\cite{rajabali2021}.

\par To conclude, we presented a back-to-back Si immersion lens setup with an asymmetric configuration allowing the spectroscopy of highly subwavelength individual THz cavities. Using this platform, we resolved the far-field transmission measurements of an ultrastrongly coupled, subwavelength split-ring single resonator to a LL transition in a single GaAs QW and a single InSb QW. The highest coupling of $60\%$ for only about $2000$ coupled electrons is reported for a single cSRR on a single InSb QW. As our results demonstrate, the combination of the aSIL configuration with a complementary-based resonant metallic structure paves the way to single-object, highly subwavelength spectroscopy of quantum electrodynamics systems operating in the mm-wave and THz range. The proposed experimental scheme  can be extended to the study of dynamical optical conductivity of  high-quality 2D structures (graphene, TMDc's, Van der Waals heterostructures)\cite{AjayanPhysToday2016} with very small effective areas (i.e. $10 \times 10$ $\mu$m$^2$) resulting from exfoliation  procedures. 

\section*{Methods}
\subsection{Asymmetric lens setup and sample fabrication} The lenses are hyperhemispherical and hemispherical ones fabricated with high resistivity Silicon (Tydex\cite{Tydex}) of diameter $2r_1=4 mm$ and $2r_2=8 mm$, respectively.
The complementary sample layout is conceived in order to ease the alignment of the front lens. The sample is metallized with a circular boundary with a diameter of $4 mm$ , matching the edge of the top lens (see Fig.\ref{fig:Schematic}c). The lens and the sample can be then accurately aligned under the optical microscope. Mechanical clamps ensure a close contact of the whole assembly front lens-sample-back lens, forming a quasi-index matched (n$_{Si}^{350GHz}=3.42$, $n_{GaAs}^{350GHz}=3.52$) stack of total length $L_s\simeq$ $6.5 mm$.  
The resonators were simulated using a commercial software package (CST microwave studio) and fabricated with a direct laser writing lithography with Heidelberg DWL66+ followed by deposition of $5 nm$ Titanium and $200 nm$ of gold and a lift-off process. 

\subsection{THz-TDS spectrometer} The measurements of the Q-factor of the cold cavity as a function of the number of resonators have been carried out with a commercial, fiber coupled THZ-TDS spectrometer (Menlo Terasmart\cite{Menlo}). THz radiation is produced and detected by a pair of InGaAs-based photoconductive antennas. The beam is guided with TPX lenses with focal length of $50 mm$ . The whole optical path is contained in a nitrogen purged box.  The measurements with the applied magnetic field are performed with a home-made THz-TDS system based on a Ti:Sapphire ultrafast ($\tau<100$ fs) laser (Mai Tai , Spectra Physics\cite{maitai}) that illuminates ($600 mW$ avg power) an interdigitated photoconductive antenna\cite{MadeoAntennaEL2010}. The THz beam is coupled to a Spectromag cryostat (Oxford Instruments\cite{Oxford}) equipped with an $11 T$ superconducting magnet and crystalline quartz (z-cut) windows . Two mirrors ($197 mm$ focal length, $2"$ diameter) focus the THz signal at the center of the superconducting coils. A variable temperature insert is used to cool down the sample at $T = 3 K$. The THz signal is then detected via electro-optic sampling in a $3 mm$ ZnTe crystal.

\newpage
\begin{addendum}

\item[Data Availability Statement]The numerical simulation and measurement data that support the plots within this paper are available from the corresponding author upon reasonable request.
\end{addendum}



\begin{addendum}

 \item[Acknowledgements] G.S. would like to thank R. Singh and F. Helmrich for discussions. We acknowledge financial support from ERC Grant No. 340975-MUSiC and the Swiss National Science Foundation (SNF) through the National Centre of Competence in Research Quantum Science and Technology (NCCR QSIT).
 \item[Competing Interests] The authors declare that they have no competing financial interests.
 \item[Authors contributions]   G. S. and S.M. conceived the aSIL configuration. S.R. fabricated the samples. S.R. and E.J. performed the measurements. M.B., C.L. and W.W. performed the epitaxial growth. Data interpretation and discussion was made by J.F., G.S., S.R, E. J. and S.M. G.S. and S.R. wrote the manuscript. All the work was made under the supervision of J.F. and G.S. 
 
 \item[Correspondence]  *Correspondence should be addressed to S. Rajabali (email: shimar@phys.ethz.ch) and G. Scalari (email: scalari@phys.ethz.ch).
\end{addendum}

\end{document}